\documentclass[12pt]{iopart}

\usepackage{booktabs}
\usepackage{threeparttable}
\usepackage{graphicx}
\usepackage{cite}

\begin{document}

\title[]{Magic wavelengths for the $6s^2\,^1S_0-6s6p\,^3P_1^o$ transition in ytterbium atom}

\author{Zhi-Ming Tang$^{1,2}$, Yan-Mei Yu$^{2}$, Jun Jiang$^{1}$ and\\ Chen-Zhong Dong$^{1}$}

\address{
$^{1}$Key Laboratory of Atomic and Molecular Physics \& Functional Materials of
Gansu Province, College of Physics and Electronic Engineering, Northwest Normal University, Lanzhou 730070, People's Republic of China\\
$^{2}$Beijing National Laboratory for Condensed Matter Physics, Institute of Physics, Chinese Academy of Sciences, Beijing 100190, People's Republic of China\\
}

\ead{ymyu@aphy.iphy.ac.cn and dongcz@nwnu.edu.cn}

\vspace{10pt}
\begin{indented}
\item[]December 2017
\end{indented}

\begin{abstract}
The static and dynamic electric-dipole polarizabilities of the $6s^2\,^1S_0$ and $6s6p\,^3P_1^o$ states of Yb are calculated by using the relativistic \emph{ab initio} method. Focusing on the red detuning region to the $6s^2\,^1S_0-6s6p\,^3P_1^o$ transition, we find two magic wavelengths at 1035.7(2) nm and 612.9(2) nm for the $6s^2\,^1S_0-6s6p\,^3P_1^o, M_J=0$ transition and three magic wavelengthes at 1517.68(6) nm, 1036.0(3) nm and 858(12) nm for the $6s^2\,^1S_0-6s6p\,^3P_1^o, M_J=\pm1$ transitions. Such magic wavelengths are of particular interest for attaining the state-insensitive cooling, trapping, and quantum manipulation of neutral Yb atom.
\end{abstract}

%
%
%
%
%

\section{Introduction} \label{sec.1}

Ytterbium has rich variety of isotopes, a $6s^2\,^1S_0$ ground state, long-lived metastable 6s6p\,$^3P^o_0$ and $^3P_2^o$ states and a number of transitions at wavelengths easily accessible by lasers for cooling and trapping. The optical trapped Ytterbium atoms provide a promising tool to study degenerate quantum gases \cite{Takasu-PRL-2004}, optical atomic clock \cite{Hinkley-Science-2013}, quantum information processing \cite{Ye-Science-2008}, and search for the CP-violating \cite{Tsigutkin-PRL-2009}, \emph{etc}. For neutral atoms, the optical trapping potentials cause spatially inhomogeneous energy shifts of the electronic states. A carefully designed optical trap with a magic wavelength that shifts the energies of the selected states equally provides a solution to this problem. As proposed by Katori \emph{et al.}, magic wavelength has been demonstrated for the atomic clock transition of Sr and Yb \cite{Katori-PRL-2003,Barber-PRL-2006}. State-insensitive trapping that is ascribed with a magic wavelength has also been enabled for Cs and Sr atoms \cite{McKeever-PRL-2003,Katori-PRL-1999,Katori-JPSJ-1999}.

The intercombination transition $6s^2\,^1S_0-6s6p\,^3P_1^o$ of Yb has a narrow linewidth around 181kHz that could cool atoms down to the photon recoil temperature of 4.4 $\mu$K \cite{Kuwamoto-PRA-1999}. In order to gain a high-density trapping, a far-off resonant trap (FORT) for Yb atom was then adopted that is overlapped on a magneto-optical trap (MOT) using the $^1S_0$ to $^3P_1^o$ transition \cite{Takasu-PRL-2003,Yamamoto-NJP-2016}. The high-density trapping and the following achievement of Bose-Einstein condensation (BEC) of $^{174}$Yb atoms have been demonstrated by using FORT, which is expected to be an important step for a future investigation of new quantum phenomena \cite{Takasu-PRL-2004}. By tuning the laser wavelength to a magic wavelength (red detuning is required for FORT), at which the $^1S_0$ and $^3P_1^o$ states have the same Ac-Stark shifts, which guarantees that the FORT is compatible with the Doppler cooling then enables high loading efficiency of magneto-optically trapped atoms. The magic wavelength in Yb and the other alkali metal atoms has been studied widely in past years \cite{Barber-PRL-2008,Lemke-PRL-2009,Brown-PRL-2017,Porsev-PRA-1999,Porsev-PRA-2004,Sahoo-PRA-2008,Dzuba-JPB-2010,Guo-JPB-2010,Safronova-PRL-2012,Beloy-PRA-2012,Porsev-PRA-2014,Wang-PRA-2016,Jiang-PRA-2016,Jiang-PRA-2017}, however most of them are concentrated on the atomic-clock transition, $^1S_0-^3P_0^o$ , whereas data of the magic wavelength of the $6s^2\,^1S_0-6s6p\,^3P_1^o$ transition is lacking, despite that the state-insensitive trapping of Yb is pursued eagerly in experiments \cite{Miranda-PhDThesis}.

In this paper, we calculate the static and dynamic electric-dipole polarizabilities of the $6s^2\,^1S_0$ and $6s6p\,^3P_1^o$ states in Yb by using the relativistic many-body calculation. We compute the magic wavelengths of the $6s^2\,^1S_0-6s6p\,^3P_1^o$ transition in Yb, considering its potential application in FORT for BEC experimental of $^{174}$Yb atoms. $^{174}$Yb has zero nuclear spin and then no hyperfine structure. We focus on the magic wavelengths that are larger than the resonant wavelength ($\lambda_R$=556nm) of the $6s^2\,^1S_0-6s6p\,^3P_1^o$ transition. We determine two magic wavelengths for the $6s^2\,^1S_0-6s6p\,^3P_1^o, M_J=0$ transition to be 612.9(2) nm and 1035.7(2) nm and three magic wavelengths for the $6s^2\,^1S_0-6s6p\,^3P_1^o, M_J=\pm1$ transition to be 1517.68(6) nm, 1036.0(3) nm, and 858(12) nm. Throughout the paper, we use atomic units (a.u.) for all energies, polarizabilities unless stated explicitly. The numerical values of the elementary charge $|e|$, the reduced Planck constant $\hbar$=$h/2\pi$, and the electron mass $m_e$ are set equal to 1 in atomic units. The atomic unit of $\alpha$ is equal to about $1.648778\times10^{-41}$C$^2$m$^2$J$^{-1}$.

\section{Method of calculation} \label{sec.2}
The ground and low lying excited states of Yb are calculated by the configuration interaction plus many-body perturbation (CI+MBPT) method that is implemented by using package~\cite{Kozlov-CPC-2015}. The computational theory and technique of the CI+MBPT method has been documented in Refs. \cite{Dzuba-PRA-1996, Dzuba-JPB-2010}. In the CI+MBPT method, the effective Hamiltonian for two valence electrons is written as
\begin{equation}
\hat{H}^{eff}=\hat{h}_1(r_1)+\hat{h}_1(r_2)+\hat{h}_2(r_1, r_2),
\end{equation}
where $\hat{h}_1$ and  $\hat{h}_2$ are the single-electron and two-electron parts of the relativistic Hamiltonian. In the implementation of the correlation operator $\hat{\Sigma}$, in addition to $\hat{\Sigma}=0$ corresponds to the standard CI method, a single-electron operator $\hat{\Sigma}_1$, representing a correlation interaction of a particular valence electron with the atomic core, and a two electron operator $\hat{\Sigma}_2$, representing screening of the Coulomb interaction between the two valence electrons by the core electrons are taken into account. In the CI+MBPT package, $\hat{\Sigma}_1$ and $\hat{\Sigma}_2$ are calculated in the second order of the MBPT. The basis set is constructed by using ``auto" generation regime that is provided in the CI+MBPT package \cite{Kozlov-CPC-2015}. The one-electron basis set includes $1s-23s$, $2p-22p$, $3d-22d$, $4f-20f$, and $5g-18g$ orbitals, where the core and $6s-9s$, $6p-9p$, $5d-8d$ orbitals are Dirac-Hartree-Fock (DHF) ones, while all the rest orbitals are virtual ones. The $V^{N-2}$ potential is used in the DHF calculation and the virtual orbitals is yielded numerically by using a recurrent relationship \cite{Kozlov-CPC-2015}.

Following the sum-over-state mythology, the dynamic polarizability $\alpha(\omega)$ is written as three parts,
\begin{equation} \label{eq:totalalpha}
\alpha(\omega)=\alpha_{v}(\omega)+\alpha_{c}(\omega)+\alpha_{vc}(\omega),
\end{equation}
where $\alpha_{v}$, $\alpha_{c}$, and $\alpha_{vc}$ represent the valence, core, and valence-core contributions, respectively, and $\omega$ is the optical frequency. When $\omega$=0, $\alpha(\omega)$ is reduced to the static polarizabilities $\alpha_0$. The $\alpha_{v}(\omega)$ part is formulated as
\begin{eqnarray}\label{eq:alphaV}
\alpha_{v}(\omega)=\alpha_{v}^{S}(\omega)
+\frac{3M_J^2-J(J+1)}{J(2J-1)}\alpha_{v}^{T}(\omega),
\end{eqnarray}
where $\alpha_{v}^{S}(\omega)$ and $\alpha_{v}^{T}(\omega)$ are the scalar and tensor polarizabilities. For a state $g$, the $\alpha_{v}^{S}(\omega)$ and $\alpha_{v}^{T}(\omega)$ can be written as
\begin{eqnarray}\label{eq:alphaVS}
\alpha_{v}^{S}(\omega)=\frac{2}{3(2J_{g}+1)}
\sum_{i}\frac{|\langle\psi_{g}\|D\|\psi_{i}\rangle|^{2}(E_{i}-E_{g})}{(E_{i}-E_{g})^{2}-\omega^{2}},
\end{eqnarray}
\begin{eqnarray}\label{eq:alphaVT}
\alpha_{v}^{T}(\omega)= & 4\left[\frac{5J_{g}(2J_{g}-1)}{6(J_{g}+1)(2J_{g}+1)(2J_{g}+3)}\right]^{1/2} \nonumber \\
& \times \sum_{i}(-1)^{J_{g}+J_{i}}\left\{
\begin{array}{ccc}
J_{g} & 1 & J_{i} \\
1 & J_{g} & 2 \\
\end{array}
\right\}
\frac{|\langle\psi_{g}\|D\|\psi_{i}\rangle|^{2}(E_{i}-E_{g})}{(E_{i}-E_{g})^{2}-\omega^{2}},
\end{eqnarray}
where $|\langle\psi_{g}\|D\|\psi_{i}\rangle|$ is the reduced matrix element of the electric dipole transition. The core-valence term is generally small, therefore it is neglected at our level of accuracy in the present work.

The core polarizability $\alpha_{c}$ is calculated by using finite-field approach. The energy of the ground state of an atom in the presence of an external weak electric field of strength $\vec{\mathcal E}$ can be expressed in the perturbation theory as  \cite{Manakov-PR-1986,Bonin-WS-1997}
\begin{eqnarray}\label{polar}
E_0(|\vec{\mathcal E}|) = E_0(0) - \frac{\alpha}{2} |\vec{\mathcal E}|^2 - \dots,
\label{eqalp}
\end{eqnarray}
where $E_0(0)$ is the energy of the state in the absence of the electric field and $\alpha$ is known as the dipole polarizability of the state. It is obvious from the above expression that $\alpha$ can be determined by evaluating the second-order differentiation of $E_0(|\vec{\mathcal E}|)$ with a small magnitude of electric field $\vec{\mathcal E}$ as
\begin{equation}
 \alpha= - \left ( \frac{ \partial^2 E_0 (|\vec{\mathcal E}|)}{\partial |\vec{\mathcal E}| \partial |\vec{\mathcal E}|} \right )_{|\vec{\mathcal E}|=0} .
\end{equation}
This procedure is known as finite-field approach for evaluating $\alpha$ which involves calculations of $E_0(|\vec{\mathcal E}|)$ after including the interaction Hamiltonian $H_{int} = -\vec{\mathcal E} \cdot \vec{D}$ with the atomic Hamiltonian. This approach has been adopted for calculation of electric dipole polarizabilities of a few atoms and ions \cite{Yu-PRA-2015,Yu-PRA-2016}. For achieving numerical stability in the result, it would be necessary to repeat the calculations by considering a number of $|\vec{\mathcal E}|$ values. This is accomplished by using the relativistic couple cluster (RCC) method that is provided in the relativistic \emph{ab initio} package DIRAC \cite{DIRAC}. We regard Yb$^{2+}$ as a closed shell system of 68 electrons. We have verified the electron correlation arising from the internal core electrons has negligible effect on $\alpha$, and therefore the atomic core $1s\cdots4d^{10}$ are frozen in the RCC calculation, while the $5s, 5p, 4f$ orbitals are correlated.  We use the Dyall's uncontracted correlated consistent double-, triple-, quadruple-$\zeta$ GTO basis sets, which are referred to as  $X\zeta$, where $X$=2, 3, and 4, respectively, \cite{Basis}. Each shell is augmented by two additional diffuse functions (d-aug) and the exponential coefficient of the augmented function is
calculated based on the following formula
\begin{equation}
\zeta_{N+1}= \left [\frac{\zeta_N}{\zeta_{N-1}} \right ]\zeta_{N} ,
\end{equation}
where $\zeta_{N}$ and $\zeta_{N-1}$ are the two most diffuse exponents for the respective atomic-shells in the original GTOs. The convergence of the results with the progressively larger basis set is checked, as presented in Table \ref{tab:Yb2+}. The final value is taken to be the $\alpha_c$ obtained by using the basis of $4\xi$. The error bar is estimated to be two times of difference of the $\alpha_c$ values obtained by using the basis of $4\xi$ and $3\xi$.

\section{Results and discussion}
Table \ref{tab:energy} presents the energies for 17 even-parity states of $6s^2$, $6s7s$, $6s8s$, $5d6s$, $6s6d$, and $6p^2$ configurations and 15 odd-parity states of $6s6p$, $6s7p$, and $6s8p$ of Yb. Our CI+MBPT results show excellent agreement with the National Institute of Standards and Technology (NIST) data for energies. Table \ref{tab:RME} presents the reduced matrix elements (RME) of the electric dipole transition among these states. While our RME values are consistent with the previously reported CI+all-order values \cite{Safronova-PRL-2012}, a big discrepancy is found in comparison of our RME value with the experimental value for the $6s^2\,^1S_0-6s6p\,^1P^o_1$ transition. The RME value for this transition has been determined experimentally to be 4.184(2)~\cite{Bowers-PRA-1996,Beloy-PRA-2012}, while our CI+MBPT and also the CI+all-order values \cite{Safronova-PRL-2012} are around 4.78-4.79, being deviated from such experimental value 16\%. This discrepancy has been noted in the previous calculations \cite{Porsev-PRA-1999,Dzuba-JPB-2010,Safronova-PRL-2012}. The reason was attributed to the missing of the $4f^{13}nln'l'n''l''$ configurations in the computation model. The importance of the $4f^{13}nln'l'n''l''$ configurations has been discussed by Dzuba, \emph{et al.}, \cite{Dzuba-JPB-2010}, which suggests that the large theory$-$experiment disagreement for the RME value of the $6s^2\,^1S_0-6s6p\,^1P^o_1$ transition is due to mixing of the $6s6p~^1P^o_1$ state with the core-excited state $4f^{13}5d6s^2\,(7/2, 5/2)_1^o$, whereas the latter is out of the computational model space of divalence system that is used in our CI+MBPT and the previous CI+all-order calculations. Since the $6s^2\,^1S_0-6s6p\,^1P^o_1$ transition contributes about 90\% to the polarizability of $6s^2\,^1S_0$, we replace the CI+MBPT value of RME for the $6s^2\,^1S_0-6s6p\,^1P^o_1$ transition by its experimental value, and include the contribution of $4f^{13}5d6s^2\,(7/2, 5/2)_1^o$ that is taken from an evaluation that was made by Beloy \cite{Beloy-PRA-2012} based on the experimental values of five lifetime results compiled in Ref. \cite{Blagoev-ADNDT-1994}. Besides, as argued by Dzuba, \emph{et al.}, such a substitution cannot be justified unless a similar correction is done for the calculation of core polarizability, $\alpha_{c}$, \emph{i.e.}, the excitation from $4f_{7/2}$ to $5d_{5/2}$ has also been included in the calculation of $\alpha_{c}$. Therefore, instead of the RPA result 6.39 \cite{Dzuba-JPB-2010}, the $\alpha_{c}$ value is taken to 7.27$\pm$0.04 in terms of our finite field calculation for the polarizability of Yb$^{2+}$. The excitation of $4f_{7/2}\rightarrow5d_{5/2}$ is taken into account in our RCC calculation. Our RCC value of $\alpha_c$ is larger than the previously reported RPA value by about 15\%.

In Table \ref{tab:alpha}, we give a breakdown of the main contributions from the intermediate states to the static polarizabilities of the $6s^2\,^1S_0$ and $6s6p\,^3P_1^o$ states. The main uncertainty is arising from the contributions of all higher lying excited states that are beyond the calculated states in our present CI+MBPT calculation that is given under ``All others". The  contribution of ``All others" is estimated based on the known knowledge of contributions of all other terms except the dominant ones, as given in the previously CI+all-order calculations \cite{Safronova-PRL-2012,Porsev-PRA-2014}. We assign the error bar ascribed with ``All others" to be 50\%. Such assignment should be reasonable because the errors of the CI+all-order values we quote and our calculated RME results both are less 5\%. Besides, through comparison of our calculated RME values with the previous CI+all-order ones \cite{Safronova-PRL-2012,Porsev-PRA-2014}, we determined a 2\% error in our calculated RMEs, while the error bar in RMEs of $6s^2\,^1S_0-6s6p\,^1P_1^o$, $6s^2\,^1S_0-6s6p\,^3P_1^o$, $6s^2\,^1S_0-4f^{13}5d6s^2\,(7/2, 5/2)_1^o$ are quoted from their references. The errors in ``All others" and our calculated RMEs are translated to the uncertainty of $\alpha^S$. Finally, the values of $\alpha^S$ for the $6s^2\,^1S_0$ and $6s6p\,^3P_1^o$ states are determined to be 135(3) and 328(19). Similarly, the tensor polarizabilities $\alpha^T$ for the $6s6p\,^3P_1^o$ state is determined be 24.1(1.5).

\begin{figure}[btp]
\centering
\includegraphics[width=9cm]{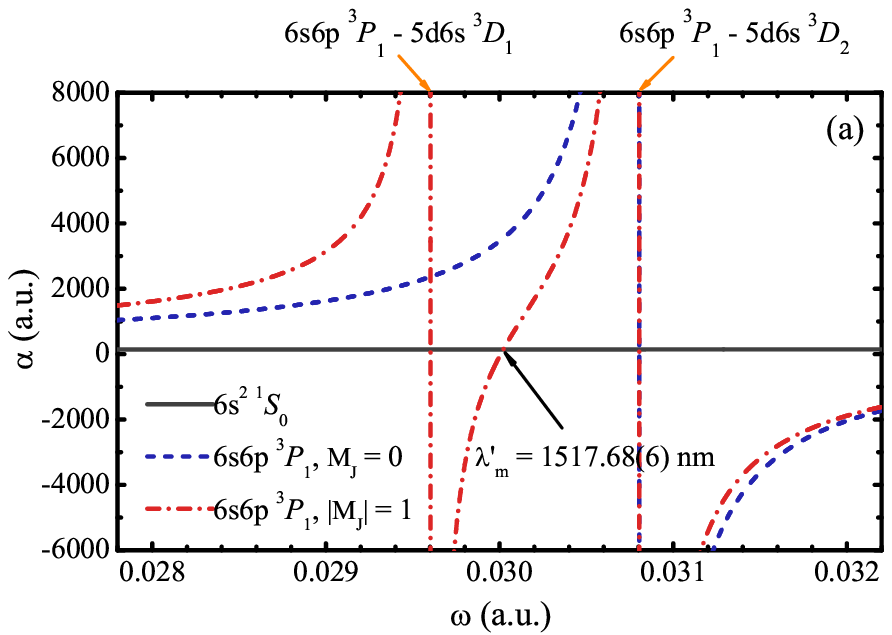}
\includegraphics[width=9cm]{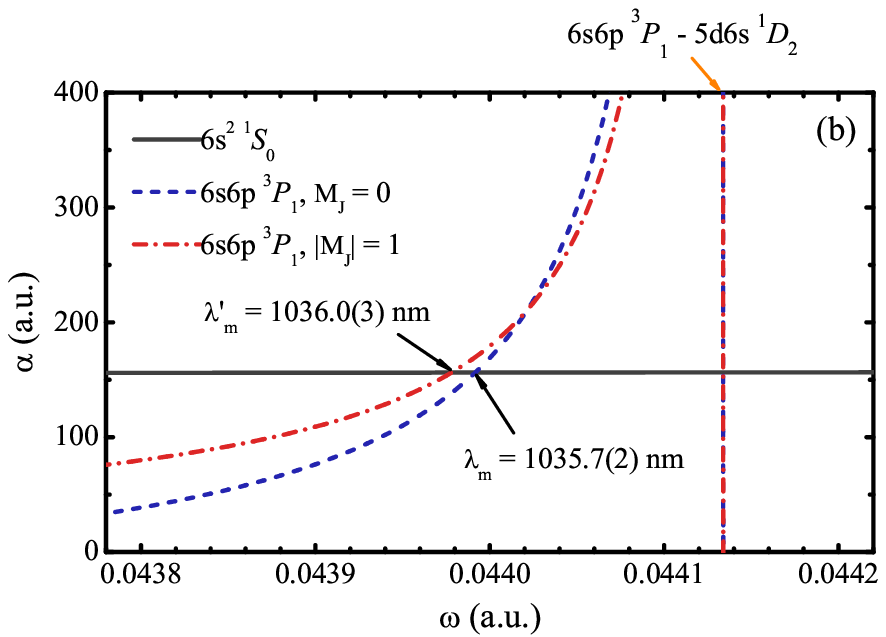}
\includegraphics[width=9cm]{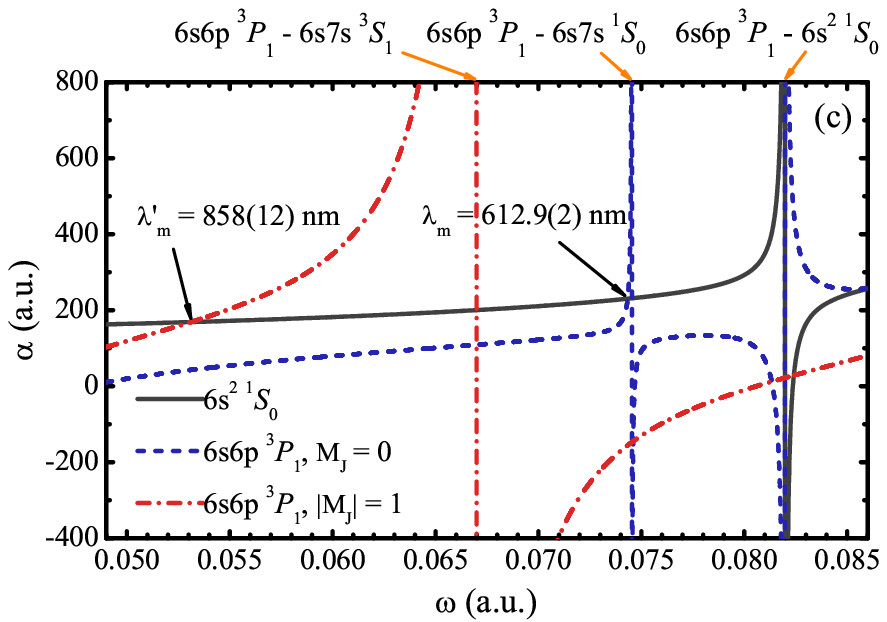}
\caption{(Color online) The frequency-dependent dynamic polarizabilities $\alpha(\omega)$ of the $6s^2\,^1S_0$ and $6s6p\,^3P_1^o$ states in Yb atom. Arrows indicate the positions of average magic wavelengths.\label{fig:alphaFre}}
\end{figure}

The dynamic polarizabilities $\alpha(\omega)$ of the $6s^2\,^1S_0$ and $6s6p\,^3P_1^o$ states are calculated in terms of Eqs. (\ref{eq:totalalpha})-(\ref{eq:alphaVT}) by summing over all the intermediate states for nonzero values of $\omega$. The ionic core polarizability depends weakly on $\omega$ and therefore is approximated by their static value. The ``All others" contributions, as considers the high lying excited states beyond the states listed in Table \ref{tab:magicB}, are taken from their static values. The role of the high lying excited states is not important for the frequencies treated here since their contributions are not resonant for the low frequencies. The total polarizability of the $6s6p\,^3P_1^o$ state depends upon its $M_J$ projection and then the magic wavelengths need to be determined separately for the cases with $M_J=0$ and $M_J=\pm1$ owing to the presence of the tensor contribution to the total polarizability of the $6s6p~^3P_1^o$ state. The magic wavelengths are found at the crossing of the $\alpha(\omega)$ curves for the $6s^2\,^1S_0$ and $6s6p\,^3P_1^o$ states. In Fig. \ref{fig:alphaFre}, we can identify two magic wavelengths for the case of $6s^2\,^1S_0-6s6p\,^3P_1^o, M_J=0$. They occur at $\lambda_m=1035.7(2)$ nm and 612.9(2) nm close to the $6s6p\,^3P_1^o-5d6s\,^3D_1$ and $6s6p\,^3P_1^o-6s7s\,^1S_0$ resonances, respectively. For the case of $6s^2\,^1S_0-6s6p\,^3P_1^o, M_J=\pm1$, three magic wavelengths, $\lambda_m'=1517.68(6)$ nm, 1036.0(3) nm, 858(12) nm are found. The first locates at a very small energy interval between the $6s6p\,^3P_1^o-5d6s\,^3D_1$ and $6s6p\,^3P_1^o-5d6s\,^3D_2$ resonances, the second locates near the $6s6p\,^3P_1^o-5d6s\,^3D_1$ resonance, and the third locates the region between the $6s6p\,^3P_1^o-5d6s\,^3D_1$ and $6s6p\,^3P_1^o-6s7s\,^1S_0$ resonances. Because the $6s6p\,^3P_1^o-5s6s\,^3D_1$ and $6s6p\,^3P_1^o-6s7s\,^3S_1$ resonances make no contribution to the total polarizabilities owing to the exact cancelation of the scalar and tensor components for the $6s6p\,^3P_1^o, M_J=0$ state, the magic wavelengths near the two resonances are absent for the $6s^2\,^1S_0-6s6p\,^3P_1^o, M_J=0$ transitions. The magic wavelength near $6s6p\,^3P_1^o-6s7s\,^1S_0$ resonance is absent for the $6s^2\,^1S_0-6s6p\,^3P_1^o, M_J=\pm1$ transitions due to the similar reason. The magic wavelength at $\lambda'_m$=858(12) nm is red detuned from the $6s^2\,^1S_0$ and $6s6p\,^3P_1^o, M_J=\pm1$ resonance with a large detuning about 302 nm that can be used for the magic wavelength ascribed with FORT. Another magic wavelength at $\lambda_m$=612.9 nm has a detuning about 57 nm.

Table \ref{tab:magicB} gives the breakdown of contributions of individual transitions to the dynamic polarizabilities at the magic wavelengths $\lambda_m$. The uncertainty in the $\lambda_m$ value is determined by variation of position of the magic wavelength due to variation of $\alpha(\omega)$ for the $6s^2\,^1S_0$ and $6s6p\,^3P_1^o$ states. The first uncertainty in $\lambda_m$ stems from error of our evaluation of ``All others" term in $\alpha(\omega)$, as denoted by Uncer.-I. The second one comes from error of RMEs used in calculation of $\alpha$. Then, the error bar is given for each magic wavelength in Table \ref{tab:magicB}. The error bar of the magic wavelength at $\lambda'_m$=858 nm is obviously larger than the other magic wavelengths, being 12 nm, and as our analysis, such large error bar is predominately caused by error in RME of $6s6p\,^3P_1^o-6s7s\,^3S_1$.

\section{Conclusion}
In summary, we have calculated the static and dynamic electric dipole polarizabilities of the $6s^2\,^1S_0$ and $6s6p\,^3P_1^o$ states by using relativistic \emph{ab initio} method. Five red-detuned magic wavelength are identified, which locate at 1035.7(2) nm and 612.9(2) nm for the case of the $6s^2\,^1S_0-6s6p\,^3P_1^o, M_J=0$ transition and 1517.68(6) nm, 1036.0(3) nm, and 858(12) nm for the $6s^2\,^1S_0-6s6p\,^3P_1^o, M_J=\pm1$ transition. Magic wavelength at $\lambda_m'$=858(12) nm is of particular interest for the far-off FORT, and the laser frequency is also readily available, for example using a Ti:sapphire laser or a tapered amplifier.

\addcontentsline{toc}{chapter}{Acknowledgment}
\section*{Acknowledgments}
The authors thank Prof. Mikhail G. Kozlov for the helpful suggestions about the use of CI-MBPT package. One of the authors, Z. M. Tang, is grateful to Dr. Jiguang Li and Dr. Zhan Bin Chen for the valuable discussions. The work was supported by the National Natural Science Foundation of China, Grants No. 91536106 and No. U1332206, and the CAS XDB21030300, and the NKRD Program of China (2016YFA0302104).

\newpage

\begin{table}[ ]
\caption{The values of electric dipole polarizabilities of Yb$^{2+}$ obtained from the finite-field approach that are implemented at the level of Dirac-Hatree-Fock (DHF) and relativistic couple cluster singles and doubles (RCCSD) calculations by using the progressively larger basis sets of Dyall's uncontracted correlated consistent double-, triple-, quadruple-$\zeta$ GTO basis sets, referred to as $2\zeta$, $3\zeta$, and $4\zeta$, respectively. \label{tab:Yb2+}}
\centering
\scriptsize
\begin{tabular}{ccc} \br
Basis  & DHF     &RCCSD              \\\mr
$2\xi$ & 6.26    &7.30                     \\
$3\xi$ & 6.36    &7.29                   \\
$4\xi$ & 6.38    &7.27                 \\ \mr
Final value & \multicolumn{2}{c}{7.27$\pm$0.04 }   \\\br
\end{tabular}
\end{table}

\begin{table}[ ]
\caption{The excited energies (in cm$^{-1}$) of the low lying excited states in Yb are obtained by using the CI+MBPT method. The absolute difference in percentage of our CI+MBPT values and the NIST energies \cite{NIST} are given in ``Diff." column. \label{tab:energy} }
\centering
\scriptsize
\begin{tabular}{llll}  \br
State               &	CI+MBPT	&	NIST   & Diff. 	\\ \mr
$6s^2\,^1S_0$	    &	0	&	0	&		  	\\
$6s6p\,^3P_0^o$	&	17446	&	17288.439	&		0.9 	\\
$6s6p\,^3P_1^o$	&	18135	&	17992.007	&		0.8 	\\
$6s6p\,^3P_2^o$	&	19858	&	19710.388	&		0.8 	\\
$5d6s\,^3D_1$	    &	25229	&	24489.102	&		3.0 	\\
$5d6s\,^3D_2$	    &	25479	&	24751.948	&		2.9 	\\
$6s6p\,^1P_1^o$	&	25698	&	25068.222	&		2.5 	\\
$5d6s\,^3D_3$	    &	25990	&	25270.902	&		2.8 	\\
$5d6s\,^1D_2$	    &	28310	&	27677.665	&		2.3 	\\
$6s7s\,^3S_1$	    &	32701	&	32694.692	&		0.02 	\\
$6s7s\,^1S_0$	    &	34287	&	34350.65	&		0.2 	\\
$6s7p\,^3P_0^o$	&	38063	&	38090.71	&		0.04 	\\
$6s7p\,^3P_1^o$	&	38133	&	38174.17	&		0.1 	\\
$6s7p\,^3P_2^o$	&	38509	&	38551.93	&		0.1 	\\
$6s6d\,^3D_1$	    &	39849	&	39808.72	&		0.1 	\\
$6s6d\,^3D_2$	    &	39882	&	39838.04	&		0.1 	\\
$6s6d\,^3D_3$	    &	40004	&	39966.09	&		0.1 	\\
$6s6d\,^1D_2$	    &	40132	&	40061.51	&		0.2 	\\
$6s7p\,^1P_1^o$	&	38894	&	40563.97	&		4.1 	\\
$6s8s\,^3S_1$   	&	41563	&	41615.04	&		0.1 	\\
$6s8s\,^1S_0$  	&	41965	&	41939.90	&		0.06 	\\
$6p^2\,^3P_0$ 	&	43183	&	42436.91	&		1.8 	\\
$6s8p\,^3P_0^o$	&	43612	&	43614.27	&		$<$0.01 	\\
$6s8p\,^3P_1^o$	&	43618	&	43659.38	&		0.1 	\\
$6p^2\,^3P_1$ 	&	44471	&	43805.42	&		1.5 	\\
$6s8p\,^3P_2^o$	&	43797	&	43806.69	&		0.02 	\\
$6s8p\,^1P_1^o$	&	43861	&	44017.60	&		0.4 	\\
$6s7d\,^3D_1$ 	&	44611	&	44311.38	&		0.7 	\\
$6s7d\,^3D_2$ 	&	44496	&	44313.05	&		0.4 	\\
$6s7d\,^1D_2$ 	&	44583	&	44357.60	&		0.5 	\\
$6s7d\,^3D_3$ 	&	44569	&	44380.82	&		0.4 	\\
$6p^2\,^3P_2$ 	&	45768	&	44760.37	&		2.3 	\\\br
\end{tabular}
\end{table}

\begin{table}[ ]
\caption{Reduced electric-dipole transition matrix elements starting from the $6s^2\,^1S_0$ and $6s6p\,^3P_1^o$ states (in a.u.) calculated by using the CI+MBPT method. Values from the CI+all-order calculation \cite{Safronova-PRL-2012} and values translated from experimental measurements of state lifetimes are given where available for comparison. \label{tab:RME} }
\centering
\scriptsize
\setlength{\tabcolsep}{2pt}
\begin{threeparttable}
\begin{tabular}{lccc} \br
 Transition	                        &  CI+MBPT  &CI+all-order                    &	Exp.                       \\ \mr
$6s^2\,^1S_0-6s6p\,^3P_1^o$	&	0.58 	&0.571~\cite{Safronova-PRL-2012} &0.543(11)~\cite{Bowers-PRA-1996} 	 	                       \\
$6s^2\,^1S_0-6s6p\,^1P_1^o$	&	4.79 	&4.78~\cite{Safronova-PRL-2012}	 &4.148(2)~\cite{Takasu-PRL-2004}  \\
$6s^2\,^1S_0-6s7p\,^3P_1^o$	&	0.08    &	            	                &	 	    \\
$6s^2\,^1S_0-6s7p\,^1P_1^o$	&	0.67 	&0.65~\cite{Safronova-PRL-2012}	 &	 	    \\
$6s^2\,^1S_0-6s8p\,^3P_1^o$	&	0.01  	&	            	                &	 	    \\
$6s^2\,^1S_0-6s8p\,^1P_1^o$	&	0.21    &	             	                &	 	    \\
$6s^2\,^1S_0-4f^{13} 5d6s^2\,(7/2,5/2)^o_1$  &                         &	        &2.04(6)$^a$ \cite{Beloy-PRA-2012}      	               	 	    \\
$6s6p\,^3P_1^o-5d6s\,^3D_1$	&	2.57 	&2.51~\cite{Porsev-PRA-2014}  &	 	    \\
$6s6p\,^3P_1^o-5d6s\,^3D_2$	&	4.45 	&4.35~\cite{Porsev-PRA-2014}  &	 		\\
$6s6p\,^3P_1^o-5d6s\,^1D_2$	&	0.46 	&0.453~\cite{Porsev-PRA-2014} &	 		\\
$6s6p\,^3P_1^o-6s6d\,^3D_1$	&	1.57 	&1.62~\cite{Porsev-PRA-2014}  &	 		\\
$6s6p\,^3P_1^o-6s6d\,^3D_2$	&	2.72 	&2.78~\cite{Porsev-PRA-2014}  &	 		\\
$6s6p\,^3P_1^o-6s6d\,^1D_2$	&	0.53 	&		 	                  &	 		\\
$6s6p\,^3P_1^o-6s7d\,^3D_1$	&	2.47 	&			                  &	 		\\
$6s6p\,^3P_1^o-6s7d\,^3D_2$	&	1.67 	&		                      &	 		\\
$6s6p\,^3P_1^o-6s7d\,^1D_2$ &	0.86 	&		                      &	 		\\
$6s6p\,^3P_1^o-6s7s\,^3S_1$	&	3.47 	&3.46~\cite{Porsev-PRA-2014}  &	 		\\
$6s6p\,^3P_1^o-6s7s\,^1S_0$	&	0.24 	&0.243~\cite{Porsev-PRA-2014} &	 		\\
$6s6p\,^3P_1^o-6s8s\,^3S_1$	&	1.00 	&		 	 &	 		\\
$6s6p\,^3P_1^o-6s8s\,^1S_0$	&	0.30 	&		     &	 		\\
$6s6p\,^3P_1^o-6p^2\,^3P_0$	&	2.59 	&		     &	 		\\
$6s6p\,^3P_1^o-6p^2\,^3P_1$	&	0.18 	&		     &	 		\\
$6s6p\,^3P_1^o-6p^2\,^3P_2$	&	2.92 	&		 	 &	 		\\ \br
\end{tabular}
\begin{tablenotes}
\item[a] This RME is derived from Beloy's evaluation who determined contribution of $4f^{13}5d6s^2\,(7/2, 5/2)_1^o$ to polarizability of $6s^2\,^1S_0$ by taking the weighted mean of five state lifetimes \cite{Beloy-PRA-2012}.
\end{tablenotes}
\end{threeparttable}
\end{table}

\begin{table}[ ]
\caption{The breakdown of the contributions of the static scalar polarizabilities $\alpha^{S}$ for the $6s^2\,^1S_0$ and $6s6p\,^3P_1^o$ states and the tensor polarizabilities $\alpha^{T}$ for the $6s6p\,^3P_1^o$ state of Yb atom. The energies are taken from NIST data. The values of RMEs are taken from our CI+MBPT calculation except three of them are substituted by the experimental values as cited. Each contribution is given under ``$\alpha_i$". ``All others" represents the contribution of all other valence states not explicitly given in the table. ``$\alpha_{core}$" represents the contribution from the core, taken from our calculation. The present results are compared with other theoretical and experimental values. Uncertainties in the last digits are given in parentheses where available.  \label{tab:alpha}}
\centering
\scriptsize
\setlength{\tabcolsep}{2pt}
\begin{threeparttable}
\begin{tabular}{lllc}
\br
Polariz.          	    &	Contrib.            &$|{\langle}g\|D\|m{\rangle i}|$ & $\alpha_i$   \\ \mr
$\alpha^{S}(^1S_0)$   &	$6s6p\,^1P_1^o$	&4.148(2)~\cite{Bowers-PRA-1996} &	100.4(1)  \\
                        &	$6s6p\,^3P_1^o$	&0.543(11)~\cite{Takasu-PRL-2004} &	2.4(1)	   \\
	                    &	$6s7p\,^1P_1^o$	&0.67 	                         &	1.62 	   \\
	                    &	$6s7p\,^3P_1^o$	&0.08 	                         &	0.021       \\
	                    &	$6s8p\,^1P_1^o$	&0.21  	                         &	0.15      \\
	                    &	$6s8p\,^3P_1^o$	&0.014     	                     &	0.0007     \\
                 &$4f^{13} 5d6s^2 (7/2,5/2)^o_1$&2.04(6)~\cite{Beloy-PRA-2012}  &21.1(1.2)  \\
	                    &	All others	        &		           &2.04$^a$              \\
	                    &	$\alpha_{core}$	    &                   &7.27(4) 		                	\\
	                    &	Total	            &                   &135(3)              \\
	                    &	Refs.               &                   &118(45)~\cite{Porsev-PRA-1999},141(6)~\cite{Dzuba-JPB-2010},                           \\
	                    &		                &                   &144.59~\cite{Sahoo-PRA-2008},139(15)~\cite{Guo-JPB-2010},  \\	
                        &                       &                   &141(2)~\cite{Safronova-PRL-2012},    \\
                        &                       &                   &$134.4\pm1.0<\alpha^{S}<144.2\pm1.0$ \cite{Beloy-PRA-2012}   \\          \mr
$\alpha^{S}(^3P_1^o)$	&	$6s^2\,^1S_0$	    &	0.58 	  	    &-0.91 							\\
	                    &	$5d6s\,^3D_1$	    &	2.57 	        &49.44  	           \\
	                    &	$5d6s\,^3D_2$	    &	4.45 	        &142.96  	             \\
	                    &	$5d6s\,^1D_2$	    &	0.46 	        &1.04            	\\
	                    &	$6s6d\,^3D_1$	    &	1.57 	        &5.54     	      \\
	                    &	$6s6d\,^3D_2$   	&	2.72 	        &16.46             	\\
	                    &	$6s6d\,^1D_2$	    &	0.53 	        &0.62  	          \\
	                    &	$6s7d\,^3D_1$	    &	2.47  	        &11.33      	  \\
	                    &	$6s7d\,^3D_2$   	&	1.67            &5.14	         \\
	                    &	$6s7d\,^1D_2$	    &	0.86  	        &1.36       	   \\
	                    &	$6s7s\,^3S_1$	    &	3.47 	        &39.91      	    \\
	                    &	$6s7s\,^1S_0$	    &	0.24 	        &0.17     	        \\
	                    &	$6s8s\,^3S_1$	    &	1.00  	        &2.06            	\\
	                    &	$6s8s\,^1S_0$   	&	0.29  	        &0.18                 		\\
                        &   $6p^2\,^3P_0$	    &	2.59  	        &13.43              		\\
                        &   $6p^2\,^3P_1$	    &	0.18  	        &0.06                  		\\
                        &   $6p^2\,^3P_2$	    &	2.92   	        &15.59                 	\\
                        &	All others	        &		            &16.87$^b$	               	\\
	                    &	$\alpha_{core}$  	&		            &7.27(4)   	               \\
	                    &	Total	            &		            &328(19)                          \\
                        &   Refs.               &                   &278(15)~\cite{Porsev-PRA-1999},315(11)~\cite{Porsev-PRA-2014}     \\\mr
$\alpha^{T}(^3P_1^o)$	&	                    &		            &24.1(1.5)          \\
                        &	Refs.               &		            &24.3(1.5)~\cite{Porsev-PRA-1999},24.06(1.37)~\cite{Rinkleff-ZPA-1980}, \\
                        &	                    &		            &24.26(84)~\cite{Kulina-ZPA-1982},23.33(52)~\cite{Li-JPB-1995} 	\\\br
\end{tabular}
\begin{tablenotes}
\item[a] This value is estimated by subtract contributions of $6s7p\,^3P_1^o$ and $6s8p$ from the value of ``All others" given in Ref. \cite{Safronova-PRL-2012}.
\item[b] This value is estimated by subtract contributions of $6s8s$, $6s7d$ and $6p^2\,^3P_J^o$ from the value of ``All others" given in Ref. \cite{Porsev-PRA-2014}.
\end{tablenotes}
\end{threeparttable}
\end{table}


\begin{table}[]
\caption{ The contributions of individual transitions to the polarizabilities (in a.u.) at the magic wavelengths, $\lambda_{m}$ and $\lambda'_{m}$, (in nm) for the $6s^2\,^1S_0-6s6p\,^3P_1^o, M_J=0$ and $6s^2\,^1S_0-6s6p\,^3P_1^o, M_J=\pm1$, respectively. The numbers in parentheses are uncertainties in the last digits.\label{tab:magicB}}
\centering
\scriptsize
\setlength{\tabcolsep}{3pt}
\begin{tabular}{lllllll} \br
                               &  \multicolumn{2}{c}{$M_J=0$}&&\multicolumn{3}{c}{$M_J=\pm1$} \\\cline{2-3} \cline{5-7}
                               &	1035.7(2) &	612.9(2)      &&	1517.68(6) &	1036.0(3) &	858(12) 	 	\\	
Uncer.-I	                   &	0.10 	&	0.10 &&	0.042 	&	0.20 	&	8.2 	    \\	
Uncer.-II	                   &	0.06 	&	0.10 &&	0.017 	&	0.09 	&	4.2 	 	\\	\mr
		\multicolumn{7}{l}{$6s^2\,^1S_0$}					\\	
$6s6p\,^1P_1^o$	&	117.918   	&	174.215   	&	&	107.879   	&	117.906   	&	128.128   	\\
$6s6p\,^3P_1^o$	&	3.36757 	&	13.4883  	&	&	2.76919 	&	3.36672 	&	4.13213 	\\
$6s7p\,^1P_1^o$	&	1.71532 	&	1.93041 	&	&	1.66200 	&	1.71526 	&	1.76378 	\\
$6s8p\,^1P_1^o$	&	0.16130 	&	0.17799 	&	&	0.15706 	&	0.16130 	&	0.16512 	\\
$6s7p\,^3P_1^o$	&	0.02285 	&	0.02616 	&	&	0.02204 	&	0.02284 	&	0.02358 	\\
$6s8p\,^3P_1^o$	&	0.00069 	&	0.00077 	&	&	0.00068 	&	0.00069 	&	0.00071 	\\
$4f^{13}5d6s^2\,(7/2,5/2)_1^o$	&	23.7609  	&	31.0141  	&	&	22.2616  	&	23.7591  	&	25.2151 	\\
All others                     &2.04 	&	2.04 	&	&	2.04 	&	2.04 	&	2.04 \\
$\alpha_{core}$	               &7.27 	&	7.27 	&	&	7.27 	&	7.27 	&	7.27 \\	
Total	                       &156 	&	230 	&	&	144 	&	156 	&	169 \\	\mr
		\multicolumn{7}{l}{$6s6p\,^3P_1^o$}				\\
$6s^2\,^1S_0$		&	-3.84505 	&	-15.4008 	&	&	0 	&	0 	&	0 	\\
$5d6s\,^3D_1$		&	0 	&	0 	&	&	-2603.08 	&	-61.4402 	&	-33.4244 	\\
$5d6s\,^3D_2$		&	-164.954 	&	-35.5567 	&	&	2576.42 	&	-123.866 	&	-65.2025 	\\
$5d6s\,^1D_2$		&	198.075 	&	-0.68009 	&	&	1.74439 	&	135.320 	&	-2.09038 	\\
$6s6d\,^3D_1$		&	0 	&	0 	&	&	9.14158 	&	10.3296 	&	11.6266 	\\
$6s6d\,^3D_2$		&	24.5451 	&	44.6557 	&	&	16.2955 	&	18.4061 	&	20.7086 	\\
$6s6d\,^1D_2$		&	0.91309 	&	1.62798 	&	&	0.60794 	&	0.68472 	&	0.76798 	\\
$6s7s\,^3S_1$		&	0 	&	0 	&	&	74.9143 	&	105.212 	&	161.165 	\\
$6s7s\,^1S_0$		&	0.80542 	&	97.6243 	&	&	0 	&	0 	&	0 	\\
$6s8s\,^3S_1$		&	0 	&	0 	&	&	3.34266 	&	3.70036 	&	4.07467 	\\
$6s8s\,^1S_0$		&	0.63413 	&	0.99097 	&	&	0 	&	0 	&	0 	\\
$6p^2\,^3P_0$		&	47.7388 	&	72.6530 	&	&	0 	&	0 	&	0 	\\
$6p^2\,^3P_1$		&	0 	&	0 	&	&	0.09332 	&	0.10142 	&	0.10959 	\\
$6p^2\,^3P_2$		&	21.5007 	&	29.7570 	&	&	14.9324 	&	16.1240 	&	17.3099 	\\
$6s7d\,^3D_1$		&	0 	&	0 	&	&	18.1288 	&	19.6331 	&	21.1388 	\\
$6s7d\,^3D_2$		&	7.12690 	&	10.0160 	&	&	4.93522 	&	5.34466 	&	5.75448 	\\
$6s7d\,^1D_2$		&	1.88309 	&	2.64228 	&	&	1.30439 	&	1.41218 	&	1.52001 	\\
All others                     &14.57 	&	14.57 	&	&	18.01 	&	18.01 	&	18.01 \\
$\alpha_{core}$                &7.27 	&	7.27 	&	&	7.27 	&	7.27 	&	7.27 \\
Total	                       &156 	&	230 	&	&	144 	&	156 	&	169 \\\br
\end{tabular}
\end{table}

\end{document}